\documentclass[fleqn,twoside]{article}
\usepackage{espcrc2}

\usepackage{graphicx}
\usepackage[figuresright]{rotating}

\newcommand{\AmS}{{\protect\the\textfont2
  A\kern-.1667em\lower.5ex\hbox{M}\kern-.125emS}}
\newcommand{\bgamma}{\mbox{\boldmath $\gamma $}}  
\newcommand{\bto}{\mbox{\boldmath $\to $}}

\title{Weak radiative hyperon decays}

\author{P. \.Zenczykowski \address[IFJ]{Institute of Nuclear Physics\\ 
        Radzikowskiego 152, 31-342 Krak\'ow, Poland}%
        \thanks{This work was supported in part by the Polish State Committee
	for Scientific Research grant 5 P03B 050 21.}}
       
\begin{document}

\begin{abstract}
The problem of weak radiative hyperon decays (WRHD) is reviewed. 
With the recent measurement of the $\Xi ^0 \to \Lambda \gamma$
asymmetry confirming Hara's theorem, implications from its violation
in low-energy theoretical approaches are discussed.
It is shown 
how an underlying symmetry link should be formulated
for a successful description of both
  nonleptonic and radiative weak hyperon decays.
The sign of the $\Xi ^0 \to \Lambda \gamma$ asymmetry 
and the overall size of parity-violating WRHD amplitudes together lead
to the resolution of 
the old S:P problem in nonleptonic decays.

\vspace{1pc}
\end{abstract}

\maketitle

\section{Introduction}

The puzzle of weak radiative hyperon decays dates back to the 60's,
when experiment \cite{exp1} suggested large asymmetry in the
 $\Sigma ^+ \to p \gamma $ decay, 
 contrary to the expectations 
 based on Hara's theorem \cite{Hara1964}. The size 
 of the asymmetry was subsequently
 confirmed, the PDG average
 now being $-0.76\pm0.08$.
 Recently new information 
 has been supplied by the NA48 measurement of
  the asymmetry of the related
 $\Xi ^0 \to \Lambda \gamma $ decay \cite{Schmidt2001}, which is
 crucial for theoretical considerations.
 
 Hara's theorem states that the
 parity-violating (p.v.) 
 amplitude of the $\Sigma ^+ \to p \gamma $ decay must
 vanish in the limit of exact SU(3). 
 If the $\Sigma ^+ \to p \gamma $
 parity-conserving (p.c.) amplitude is
 not small as the size of the branching ratio 
 suggests, the asymmetry should deviate from zero due to SU(3)
 breaking. Since the latter is usually not large,
 a small asymmetry of $\pm 0.2$ was expected.  
 The problem
  was confounded by the violation of Hara's theorem in
 the quark model and other calculations.  Since the only explicit assumptions of
 the theorem are  
 CP- and gauge-invariance,
 the origin of the latter results proved hard to understand at hadron
 level.
 With the quark model suggesting violation of the theorem,
 the question seemed to be whether large $\Sigma ^+ \to p \gamma $
 asymmetry is a sign of a strong SU(3) breaking or that of a true violation 
 of Hara's theorem.
  
 The measurement of the $\Xi ^- \to \Sigma ^- \gamma $ branching ratio
 provided the first piece of information to guide the theory:
 the observed relative size of the 
 $\Sigma ^+ \to p \gamma $ ~($suu \to duu \gamma $) and 
 $\Xi ^- \to \Sigma ^- \gamma $ ~($ssd \to sdd \gamma $)
 branching ratios excludes the possibility that single quark
   transition $s \to d \gamma $ 
 is dominant. The main contribution to the $\Sigma ^+ \to p \gamma $
 decay has to come from the two-quark ($W$-exchange) process
 $su \to ud \gamma $, as the single-quark term is 
 100 times weaker.

\section{Approaches with built-in Hara's theorem}
 The standard
 approach to WRHD was proposed by Gavela et al \cite{Gavela81},
 who presented a pole model analogous to their model
 of nonleptonic hyperon decays (NLHD) \cite{LeYaouanc79}.
 In the latter paper the
  p.v. NLHD amplitudes were given
  in terms of the current algebra (CA) commutator
 plus a correction term originating from the excited  
 negative parity $1/2^-$ baryons in intermediate states 
 (parallelling the
 dominance of the ground-state baryons in the
 p.c. amplitudes, the  $1/2^-$ baryons were
 considered here the most important contribution). 
 In \cite{Gavela81} WRHD amplitudes were described in an analogous
 way
 (the p.c. amplitudes specified by the pole model
 with intermediate ground-state baryons, while the p.v.
 amplitudes calculated from the model with intermediate
 negative parity ($1/2 ^-$) baryons). 
  Parity-violating transitions between the ground-state and the $1/2 ^-$
 baryons were driven by  $W$-exchange only.
 With SU(3) broken, ref.\cite{Gavela81} obtained a large
 asymmetry 
 \begin{equation}
 \label{Gav1}
 \alpha (\Sigma ^+ \to p \gamma ) = -0.80 ^{+0.32}_{-0.19}.
 \end{equation}
 Among other results of \cite{Gavela81}, the most important one 
 is the prediction of the $\Xi ^0 \to \Lambda \gamma $ asymmetry:
 \begin{equation}
 \label{Gav2}
 \alpha (\Xi ^0 \to \Lambda \gamma )= - 0.78
 \end{equation}

Calculations using the chiral perturbation
theory 
proved to have little predictive power \cite{Neufeld}.
This conclusion was corroborated recently by Borasoy and Holstein,
who had to discard their ChPT approach and adopt
a pole model similar to ref.\cite{Gavela81}.
Their approach (see eg. \cite{Z2000}) yields
\begin{eqnarray}
\nonumber
\alpha (\Sigma ^+ \to p \gamma )&=& -0.49 \\
\label{BH}
\alpha (\Xi ^0 \to \Lambda \gamma )&=& -1.0
\end{eqnarray}
when the singlet assignment of the $\Lambda (1405) $ intermediate state 
is used. Note the size and
the negative sign of the $\Xi ^0 \to \Lambda \gamma $ asymmetry.

Results of the application of the QCD sum rules \cite{Kha,Bal} 
were not conclusive (see Table 1),
with no prediction of \cite{Bal} for the $\Xi ^0 \to \Lambda \gamma$ 
asymmetry.\\
\phantom{x}

\noindent
Table 1\\
Predictions of QCD sum rules\\
\begin{tabular}{ccc}
\hline
asymmetry
&ref.\cite{Kha} & 
ref.\cite{Bal} \\
\hline
$\alpha (\Sigma ^+ \to p \gamma)$&
$+0.8$&${-0.85 \pm 0.15} $\\
$ \alpha (\Xi ^0 \to \Lambda \gamma)$&$
+0.9$& $---$
\end{tabular}

\section{Quark model and vector-meson dominance}
In 1983 Kamal and Riazuddin (KR) found that in the constitutent quark
model (CQM) the $W$-exchange diagrams produce a p.v.
$\Sigma ^+ \to p \gamma$ amplitude which does not vanish in the limit
of exact SU(3) \cite{KR}. 
The KR calculation has been often dismissed as erroneous, possibly
violating gauge invariance. Independent checks have confirmed 
its technical correctness, however.
The question that remained unanswered was what conclusions should be drawn
from the KR result.

An experimental way of deciding whether the CQM result constitutes
an artefact of the 
model or a real effect was pointed out in ref.\cite{LZ}.
It was observed that in
approaches violating Hara's theorem a
large and positive 
$\Xi ^0 \to \Lambda \gamma $ asymmetry is predicted, in marked contrast
to the Hara's-theorem-satisfying case.
In particular, 
in the phenomenological extension of the KR paper, 
 Verma and Sharma \cite{VS1988} obtained:
\begin{eqnarray}
\nonumber
\alpha (\Sigma ^+ \to p \gamma) &
=& -0.56\\
\label{VS}
\alpha (\Xi ^0 \to \Lambda \gamma ) &
=& {+0.68}
\end{eqnarray}

In an attempt to avoid the concepts of the CQM, ref.\cite{Zen1989} 
proposed a hadron-level
approach in which spin-flavour symmetries 
are combined with the idea of vector meson dominance (VMD).
The approach was based 
on the SU(6)-based extension of NLHD amplitudes to describe
vector-meson p.v. couplings to nucleons, as proposed
by Desplanques, Donoghue, and Holstein (DDH) \cite{DDH} for the 
description of nuclear parity violation.
It turned out that the approach of \cite{Zen1989} (here dubbed 
DDH+VMD)
leads to the violation
of Hara's theorem as well.
The origin of this result is readily traced:
the p.v. coupling of vector meson to nucleon in DDH
is of the form $\bar{u}\gamma _{\mu}\gamma _5 u V^{\mu }$. When
combined with VMD, the
 coupling  $\bar{u}\gamma _{\mu}\gamma _5 u A^{\mu }$
is generated.
As the latter coupling is not gauge-invariant by itself, violation of Hara's
theorem in the DDH+VMD approach
was blamed on gauge noninvariance of VMD.
 The issue of gauge invariance of VMD is more subtle, however. 
 In 1967 Kroll, Lee, and 
Zumino proposed that the VMD prescription consitutes an approximation
to a fully gauge-invariant quark-level contribution (KLZ) \cite{KLZ}.
Following their ideas, ref.\cite{Zen1989} accepted 
that the quark model and VMD are 
essentially equivalent gauge-invariant approaches, and that
the CQM result of \cite{KR} should
be viewed as complementary to that of DDH+VMD.
The latter approach predicted \cite{LZ}:
\begin{eqnarray}
\nonumber
 \alpha (\Sigma ^+ \to p \gamma )&
 =&  -0.95 \\
 \label{DDHVMD}
 \alpha (\Xi ^0 \to \Lambda \gamma )&
 =& { +0.8 }
\end{eqnarray}
Similarity of  Eqs. 
(\ref{VS},\ref{DDHVMD}) 
is striking. 
None\-theless, violation of Hara's theorem in CQM/DDH+VMD approaches
was still often blamed on gauge-noninvariance.

In order to exhibit 
gauge invariance of the CQM calculations, and to clarify the possible physical
origin of the violation of Hara's theorem, 
CQM calculations were recently performed \cite{Zen2001} 
in a different way, keeping  
gauge-invariance 
manifest till the very
end. Moreover,
CQM descriptions of NLHD and WRHD were discussed alongside
current algebra and VMD. The calculation
was deliberately formulated in a way
fully analogous to the standard CQM
calculations of baryon magnetic moments.
Its first step consisted in the evaluation of the modifications of
the three-quark states by the perturbing p.v.
hamiltonian
$H^{p.v.}$, which admixes 
{$q\bar{q}$} pairs into  
{$qqq$} baryon states:
\begin{eqnarray}
{H^{p.v.}}|qqq\rangle &= &
|qq{q\bar{q}}q\rangle
\end{eqnarray}
so that the $qqq\bar{q}q$ admixture has parity $P=-1$.

Thus, evaluation of photon (vector-meson) interaction with quarks
in a baryon in the presence of weak interaction
proceeds according to the standard CQM rules 
by sandwiching the interaction with photon ($A_{\mu}\to V_{\mu}$ for
vector meson) 
between appropriate quark states (Fig. 1):
\begin{eqnarray} 
\nonumber
\langle qqq|
\bar{q}\gamma _{\mu} q A^{\mu } 
|qq{q\bar{q}}q\rangle
&\phantom{xxxxxx}&\\
\langle qq{q\bar{q}}q| \bar{q}\gamma _{\mu} q A^{\mu }|qqq\rangle&&
\end{eqnarray}
The above prescription is manifestly gauge-invariant
and shows that photon and vector-meson couplings to baryons should
be proportional to each other.
Calculation performed along these lines demonstrates once again
that 
in the CQM the
 p.v. $\Sigma ^+ \to p \gamma $ amplitude is nonzero in the SU(3) limit.
At the same time, the
symmetries of the CA commutator are reproduced 
with standard pseudoscalar interaction
{$\bar{q}\gamma _5 q P$}.

\section{$\bf {\Xi ^0 \bto \Lambda \bgamma }$}
With the old value of
$\alpha (\Xi ^0 \to \Lambda \gamma)$ being $  +0.43 \pm 0.44$
as measured by \cite{James},
and the models predicting this asymmetry to be
$-0.8\pm 0.2 $ for Hara's theorem satisfied and
$+0.8 \pm 0.2$ for Hara's theorem violated, the latter
possibility was discussed
at length by the author. Since
violation of Hara's theorem in the CQM cannot be blamed
on the lack of gauge invariance 
\cite{Zen2001}, its origin
must be different. The latter was identified as 
genuine 
nonlocality of the p.v. photon-baryon
coupling in the CQM \cite{Zen2001}.
In essence, violation of Hara's theorem appears directly
connected to the old quark model question of why 
magnetic moments of quarks can be added as if the quarks were free and hence
completely independent particles which might be infinitely far apart
from each other.

\setlength{\unitlength}{0.3pt}

\begin{picture}(850,940)
\put(120,0){
\begin{picture}(700,910)
\put(0,470){
\begin{picture}(400,420)
\put(300,400){\vector(-1,0){150}}
\put(150,400){\line(-1,0){150}}
\put(300,320){\vector(-1,0){100}}
\put(200,320){\vector(-1,0){150}}
\put(50,320){\line(-1,0){50}}
\put(300,240){\vector(-1,0){100}}
\put(200,240){\line(-1,0){100}}
\put(100,240){\vector(2,-1){50}}
\put(150,215){\line(2,-1){50}}
\put(200,190){\vector(-1,0){100}}
\put(0,190){\line(1,0){100}}
\multiput(200,190)(-46,-23){4}{\line(-2,-1){36}}
\put(32,106){\vector(-2,-1){20}}
\multiput(100,240)(0,10.8){8}{\line(0,1){5}}
\multiput(0,215)(15,0){21}{\circle*{1}}
\put(10,335){\makebox(0,0){$u$}}
\put(196.8,215){\makebox(0,0){$\bar{d}$}}
\put(288,335){\makebox(0,0){$s$}}
\put(288,255){\makebox(0,0){$u$}}
\put(100,100){\makebox(0,0){$P,V,\gamma$}}
\put(360,285){\makebox(0,0){{\Large $H_u^{s,u\bar{d}}$}}}
\put(150,40){\makebox(0,0){\large $(b1)$}}
\end{picture}}
\put(0,0){
\begin{picture}(400,420)
\put(300,400){\vector(-1,0){150}}
\put(150,400){\line(-1,0){150}}
\put(300,320){\vector(-1,0){50}}
\put(250,320){\vector(-1,0){150}}
\put(100,320){\line(-1,0){100}}
\put(200,240){\vector(-1,0){100}}
\put(100,240){\line(-1,0){100}}
\put(100,190){\vector(2,1){50}}
\put(150,215){\line(2,1){50}}
\multiput(94,187)(-46,-23){2}{\line(-2,-1){36}}
\put(16,148){\vector(-2,-1){20}}
\put(200,190){\line(-1,0){100}}
\put(300,190){\vector(-1,0){100}}
\multiput(0,215)(15,0){21}{\circle*{1}}
\put(10,335){\makebox(0,0){$u$}}
\put(288,335){\makebox(0,0){$s$}}
\put(10,255){\makebox(0,0){$d$}}
\put(110,215){\makebox(0,0){$\bar{u}$}}
\multiput(200,240)(0,10.8){8}{\line(0,1){5}}
\put(80,140){\makebox(0,0){$P,V,\gamma$}}
\put(360,275){\makebox(0,0){{\Large $H^s_{u,d\bar{u}}$}}}
\put(150,70){\makebox(0,0){\large $(b2)$}}
\end{picture}
}
\end{picture}}
\end{picture}

\noindent
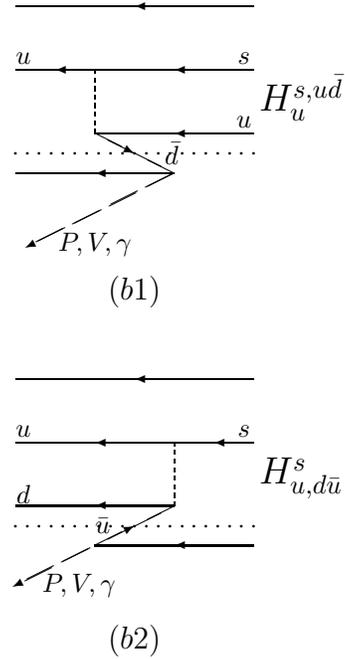
Fig.~1 Photon (meson) emission in CQM with $W$-exchange-induced admixtures\\
\phantom{x}

The p.v. amplitudes of  the $\Xi ^0 \to \Lambda \gamma $
and $\Xi ^0 \to \Sigma ^0 \gamma $ decays each
receive nonzero contributions from a different
type of diagram (see Table 2, where relevant SU(6) factors $b_{1(2)}$
are shown). Thus, the relative sign
between the contributions from (b1) and (b2) in Fig. 1
can be measured by comparing the asymmetries of these two decays.

This relative sign determines whether in $\Sigma ^+ \to p \gamma $
the contributions from diagrams (b1) and (b2) add, or completely
cancel (Table 2). Detailed considerations show
that Hara's theorem is satisfied (violated), if
$\alpha _{exp}(\Xi ^0 \to \Lambda \gamma)$
is negative (positive).

The new NA48 result \cite{Schmidt2001} of
\begin{equation}
\alpha (\Xi ^0 \to \Lambda \gamma) = -0.65 \pm 0.19 
\end{equation}
permits the following 
conclusions:

1. Hara's theorem is satisfied

2. theoretical arguments against Hara's theorem are invalid.\\
\phantom{x}

\noindent
Table 2\\
Theoretical SU(6) factors $b_{1(2)}$ for p.v. amplitudes (b1) and (b2) and experimental
asymmetries for selected WRHD (using ref.\cite{Schmidt2001})\\
\begin{footnotesize}
\begin{tabular}{rrrr}
\hline
\makebox[0in]{\phantom{\LARGE X}}
decay & $\Sigma ^+ \to p \gamma $
&$\Xi ^0 \to \Lambda \gamma $ & $\Xi ^0 \to \Sigma ^0 \gamma$\\
\makebox[0in]{\phantom{\LARGE I}}
asym. & $-0.76 {\tiny \pm 0.08} $&$-0.65 {\tiny \pm 0.19} $& $-0.63 {\tiny \pm 0.09}$\\
\hline
\makebox[0in]{\phantom{\LARGE j}}
{ $+b_1{\tiny +b_2}$}  &$ -\frac{1}{3\sqrt{3}}{\tiny -\frac{1}{3\sqrt{3}}}$
&$0 -\frac{1}{3\sqrt{3}}$&  
${+\frac{1}{3}+0 }$\\
\makebox[0in]{\phantom{\LARGE $\frac{1}{j}$}}
{ $-b_1{\tiny +b_2}$}&$ +\frac{1}{3\sqrt{3}}{\tiny -\frac{1}{3\sqrt{3}}}
$&$0 -\frac{1}{3\sqrt{3}}$& 
${-\frac{1}{3}+0}$\\
\hline
\end{tabular}
\end{footnotesize}
\phantom{x}

Consequently, CQM calculations are unphysical: while
hadron-level spin-flavour symmetries of the CQM are correct for
individual amplitudes (b1) and (b2), the CQM
connection between the latter two
is not.
Moreover, since VMD+DDH also leads to the violation of Hara's theorem,
either VMD is not universal
or the DDH approach is not fully correct.

\section{Proposed resolution}

The p.v. NLHD amplitudes can be expressed either as sums of contributions 
from the CA commutator and a correction term proportional to pion momentum 
$q$, or as sums of amplitudes
corresponding to single-quark and two-quark diagrams ($c$- and $b$- type
respectively):
\begin{eqnarray}
\nonumber
A_{NL}&=& {\rm comm.}+ R_{\mu}q^{\mu}  \\
&=&A_{NL}(b) + A_{NL}(c)
\end{eqnarray}

Symmetry considerations require that the contribution from the commutator be
proportional to the sum of SU(6) factors $b_1$ and $b_2$,
 while the correction term
be proportional to their difference:
\begin{eqnarray}
\nonumber
A_{NL}(b)&=&(b_1+b_2)b_{com}+(-b_1+b_2)b_{cor}\\
&=&b_2(b_{com}+b_{cor})\equiv b_2b_{NL}
\end{eqnarray}
The second line follows since for NLHD all $b_1$'s  are zero.
The NLHD data fix the two parameters 
defining the size of $b$- and $c$- 
amplitudes as:
\begin{eqnarray}
\nonumber
b_{NL}=&-5\cdot 10^{-7}&\\
\label{NLexp}
c_{NL}=&+12 \cdot 10^{-7}&
\end{eqnarray}
($A_{NL}(c)$ is proportional to $c_{NL}$),
or equivalently,
\begin{eqnarray}
d_S&=&b_{NL}\\
f_S&=&-b_{NL}+2c_{NL}/3
\end{eqnarray}
corresponding to the $f_S/d_S=-2.6$ ratio for the S-waves.
However, precisely because all $b_1$'s are zero, these data
 do not allow one to fix the size of $b_{com}$ and $b_{cor}$
separately.

For WRHD the contribution from single-quark diagrams is  negligible.
Furthermore, there can be no $b_1+b_2$ terms (Hara's theorem is satisfied).
Thus, the WRHD amplitudes are proportional to the
differences
 of SU(6) factors $b_1$ and $b_2$:
 \begin{equation}
 A_{WR}=(-b_1+b_2)b_{WR}\cdot e/g
 \end{equation}
where 
factor $e/g$ converts between strong ($g$) 
and electromagnetic ($e$) couplings,  so that 
the spin-flavor symmetry link for the $-b_1+b_2$ terms in NLHD and WRHD
has the simple form:
\begin{equation}
\label{corWR}
b_{cor}=b_{WR}
\end{equation}
Previous successful description of 
the branching ratios of $\Xi ^0 \to \Lambda \gamma$
and $\Xi ^0 \to \Sigma ^0 \gamma $ decays (see \cite{LZ}) indicates that,
numerically,
\begin{equation}
|b_{WR}|\approx 5 \cdot 10^{-7} = |b_{NL}|
\end{equation}

If $b_{WR}\approx b_{NL}$, Eqs. (\ref{NLexp},\ref{corWR}) imply that
$b_{com}=0$, which follows from CQM if Hara's theorem
is to be satisfied \cite{Zen2001}.
Since in this case positive $\Xi ^0 \to \Sigma ^0 (\Lambda) \gamma$ 
asymmetries are predicted  in disagreement with
experiment, this possibility has to be rejected.
The only other possibility is 
$b_{WR}\approx -b_{NL}$.
 This leads to correct 
 negative $\Xi ^0 \to \Sigma ^0 (\Lambda) \gamma$ 
asymmetries.
One concludes that
$b_{com}=2b_{NL}$.

Single-quark contributions to the NLHD amplitudes obtained by symmetry
from WRHD should be negligible, ie. $c_{NL}=c_{com}$.
The $f/d$ ratios for the S- and P- waves of NLHD amplitudes appear therefore
different, since
for the P-waves one obtains
\begin{eqnarray}
d_P=&b_{com}  &\approx 2b_{NL}\\
\nonumber
f_P=&-b_{com}+2c_{com}/3&\approx -2b_{NL}+2c_{NL}/3
\end{eqnarray}
leading to a 
resolution of the S:P problem in NLHD:
\begin{eqnarray}
\nonumber
d_P/d_S &\approx &2 \\
\nonumber
f_P/f_S &\approx &1.4\\
f_P/d_P &\approx &-1+c_{NL}/(3b_{NL})=-1.8
\end{eqnarray}

In Table 3 
we compare the WRHD data with the predictions
of Gavela et al \cite{Gavela81}. 
The last column gives the branching ratios and asymmetries
for the decays of $\Xi ^0$ in an SU(6) symmetric 
approach just discussed. The relevant entries
are obtained from those given in \cite{LZ} by just reversing the sign
of the $\Xi ^0 \to \Lambda \gamma $ amplitude.
(Description of $\Sigma ^+$ and $\Lambda $ decays requires 
inclusion of SU(3) breaking and more modifications
than a simple
sign reversion).\\
\phantom{x}

\noindent
Table 3\\
Comparison of branching ratios and asymmetries\\
\phantom{x}

\noindent
Branching ratios\\
\begin{footnotesize}
\begin{tabular}{rrrr}
\hline
 decay   & exp & \cite{Gavela81} & \cite{LZ} modif.\\
\hline
$\Sigma ^+ \to p \gamma$            
& $\phantom{-}1.23\pm0.06$ & $\phantom{-}0.92^{+0.26}_{-0.14}$ 
&  {\rm not simple} \\
$\Lambda \to n \gamma$ & $1.75\pm0.15$ & $ 0.62$  & {\rm not simple} \\
$\Xi ^0 \to \Lambda \gamma $& $1.06\pm 0.16$ & $3.0$  & $0.9-1.0$   \\
$\Xi ^0 \to \Sigma ^0 \gamma$ & $3.56 \pm 0.43 $ & $7.2$ & $4.0-4.1$  \\
\hline
\end{tabular}
\end{footnotesize}\\
\noindent
Asymmetries\\
\begin{footnotesize}
\begin{tabular}{rrrr}
\hline
 decay   & exp & \cite{Gavela81} & \cite{LZ} modif.\\
\hline
$ \Sigma ^+ \to p \gamma $ & $-0.76 \pm 0.08$ & $-0.80^{+0.32}_{-0.19}$ 
& {\rm not simple} \\
$\Lambda \to n \gamma $ & & $-0.49$ & {\rm not simple}  \\
$\Xi ^0 \to \Lambda \gamma $&$-0.65 \pm 0.19$&$-0.78$& $-0.8$\\
$\Xi ^0 \to \Sigma ^0 \gamma $&$-0.65 \pm 0.13$&$-0.96$&$-0.45$\\
\hline
\end{tabular}
\end{footnotesize}

\section{Conclusions}

The new NA48 result 
confirms Hara's theorem.
Consequently,
large SU(3) breaking  is needed to describe the
$\Sigma ^+ \to p \gamma $ asymmetry.
Furthermore, the CQM result appears as an artefact of the model.
Thus, the CQM constitutes an abstraction from spin-flavor symmetries
of hadronic amplitudes that goes too far.
Since Hara's theorem violation was also predicted in a symmetry-based
framework which combined current-field identity with an approach
describing p.v. couplings of vector mesons to nucleons 
(used in explanations of nuclear parity violation \cite{Despl}), 
it follows that
either current-field identity is not universal, or
our present understanding of nuclear parity violation is not fully correct.

The proposed resolution of the problem of NLHD-WRHD symmetry connection 
implies that symmetry should be imposed 
at the level of axial/vector currents (and not between the couplings of
pseudoscalar and vector fields to baryons if VMD is to hold) \cite{GM1964}.
When this is done, the old
S:P problem in NLHD is explained automatically.

\end{document}